\documentclass[english,aps,preprint]{revtex4}
\usepackage[T1]{fontenc}
\usepackage[latin9]{inputenc}
\usepackage{graphicx}

\makeatletter
\@ifundefined{textcolor}{}
{%
 \definecolor{BLACK}{gray}{0}
 \definecolor{WHITE}{gray}{1}
 \definecolor{RED}{rgb}{1,0,0}
 \definecolor{GREEN}{rgb}{0,1,0}
 \definecolor{BLUE}{rgb}{0,0,1}
 \definecolor{CYAN}{cmyk}{1,0,0,0}
 \definecolor{MAGENTA}{cmyk}{0,1,0,0}
 \definecolor{YELLOW}{cmyk}{0,0,1,0}
}

\usepackage{babel}

\makeatother

\usepackage{babel}
\begin{document}

\title{Probing Anomalous $HZZ$ Couplings at the LHeC}

\author{I. T. Cakir}

\email{ilkay.turkcakir@gmail.com}

\selectlanguage{english}%

\affiliation{Ankara University, Department of Physics, 06100, Ankara, Turkey}

\author{O. Cakir}

\email{ocakir@science.ankara.edu.tr}

\selectlanguage{english}%

\affiliation{Ankara University, Department of Physics, 06100, Ankara, Turkey}

\author{A. Senol}

\email{asenol@kastamonu.edu.tr}

\selectlanguage{english}%

\affiliation{Kastamonu University, Department of Physics, 37100, Kastamonu, Turkey }

\affiliation{Abant Izzet Baysal University, Department of Physics, 14280, Bolu,
Turkey}

\author{A. T. Tasci}

\email{atasci@kastamonu.edu.tr}

\selectlanguage{english}%

\affiliation{Kastamonu University, Department of Physics, 37100, Kastamonu, Turkey}
\begin{abstract}
We examine the sensitivity to the couplings of the Higgs boson to
neutral gauge bosons in a model independent way at the Large Hadron
electron Collider (LHeC). We have obtained the constraints on anomalous
couplings for $HZZ$ vertex via the process $e^{-}p\rightarrow e^{-}HqX$.
We find the accessible limits of the anomalous coupling $b_{Z}$ as
$(-0.12,0.43)$ and $(-0.10,0.33)$, while the limits on coupling
$\beta_{Z}$ as ($-0.32,0.32$) and $(-0.24,0.24)$ at the electron
beam energy $E_{e}=60$ GeV and $E_{e}=140$ GeV, respectively.
\end{abstract}


\maketitle

\section{Introduction}

The recent discovery of a new boson with a mass of 125 GeV at the
CERN LHC \cite{ATLAS2012,CMS2012} matches many of the properties
of the Higgs boson from the Higgs mechanism within the standard model
(SM). With more data analyzed, the LHC is expected to be able to probe
the detailed properties of this new scalar particle, and verify its
nature.

A first determination of the couplings, spin and parity properties
can be performed with the current results. The ATLAS \cite{ATLAS2013}
and CMS \cite{CMS2013} Collaborations have presented that the observed
state resembles very closely to the Higgs boson. Searching for the
properties of this new boson can extend our knowledge of the electroweak
symmetry breaking (EWSB) sector. This search has also very important
consequences on theories beyond the SM.

From the theoretical side, there could be different EWSB scenarios
in which the Higgs boson can be elementary and weakly interacting
\cite{Altarelli2012} or composite and related to a new strongly interacting
sector \cite{Dimopoulos1979}. Under the assumption there is new physics
associated with the EWSB sector, its effects on the phenomenology
of the Higgs boson can be parametrized in terms of an effective Lagrangian
at the electroweak scale. In this framework, the corrections from
higher dimensional operators to the Higgs boson $(H$) couplings to
the gauge bosons ($V$) can be expressed through an effective Lagrangian.
The efffective operators give rise to anomalous $HVV$ couplings,
and these operators can modify both the Higgs boson production and
its decay rates \cite{Corbett2012}. 

Extensive studies about the anomalous $HZZ$ couplings described by
a model independent effective Lagrangian approach have been performed
in the literature at hadron colliders \cite{Corbett2012,Eboli1998,Gonzalez1999,Eboli2000,Zhang2003,Bonnet2012,Christense2010,Hankele2006}
and lepton colliders \cite{Biswal2006,Hagivara1993,Hagiwara1994,Grzadkowski1995,Gounaris1996,He2003,Barger2003,Han2006,Choudhury2006 ,Dutta2008,Rindatti2009}.

The predictions on the measurement of ratios of the Higgs couplings
with an accuracy at the level of a few percent have been given in
\cite{Djouadi2013} for the LHC at $\sqrt{s}=14$ TeV with a high
luminosity of 300 fb$^{-1}$. Furthermore, the precision measurement
of the self-couplings of this new boson is very important for the
future experiments since it could well point the new physics beyond
the SM.

The LHeC could probe the $HZZ$ couplings without any assumption on
the $HWW$ couplings. It has the uniqe oppurtunity to probe $HZZ$
couplings in Higgs production via weak vector boson fusion, in contrast
to the production at the LHC where the contributions come from the
$HWW$ and $HZZ$ couplings. Therefore, it is a particular interest
since these couplings could receive sizeable anomalous contributions
from physics beyond the SM. 

A high energy electron-proton collider can be realised by accelerating
electrons in a linear accelerator (linac) to 60-140 GeV and colliding
them with the 7 TeV protons incoming from the LHC. The anticipated
integrated luminosity is about in order of $10$ and $100$ $fb^{-1}$
\cite{LHeC2012}. 

In this work, we study the sensitivity to the anomalous $HZZ$ couplings
in a model independent way at the Large Hadron electron Collider (LHeC)
via the production process $e^{-}p\rightarrow e^{-}HqX$. We calculate
the production cross sections depending on the anomalous couplings.
The results of our analyses are given through the $95$\% C.L. contour
plots.

\section{The HZZ Couplings}

Providing the Lorentz and gauge invariance, a general $HZZ$ couplings
structure may be expressed \cite{Hagiwara1994,Hagivara2000} as 

\begin{equation}
\Gamma_{\mu\nu}=g_{Z}\left[a_{Z}g_{\mu\nu}+\frac{b_{Z}}{m_{Z}^{2}}(k_{2\mu}k_{1\nu}-g_{\mu\nu}k_{1}.k_{2})+\frac{\beta_{Z}}{m_{Z}^{2}}\epsilon_{\mu\nu\alpha\beta}k_{1}^{\alpha}k_{2}^{\beta}\right]\label{eq:1}
\end{equation}
where $k_{1}$ and $k_{2}$ are the momenta of two $Z$ bosons with
$g_{Z}=2g_{e}m_{Z}/sin2\theta_{W}.$ In the context of the SM, at
the tree level, $a_{Z}=1$, while the other couplings $b_{Z}$ and
$\beta_{Z}$ vanish identically. We define $\Delta a_{Z}=a_{Z}-1$
in order to deal with a SM like Higgs boson, and hence we set $\Delta a_{Z}=0$
in the SM case. The couplings $b_{Z}$ and $\beta_{Z}$ can arise
from higher order terms in an effective theory, where $b_{Z}$ and
$\beta_{Z}$ are the $CP$ conserving and $CP$ violating couplings,
respectively.

\begin{figure}
\includegraphics{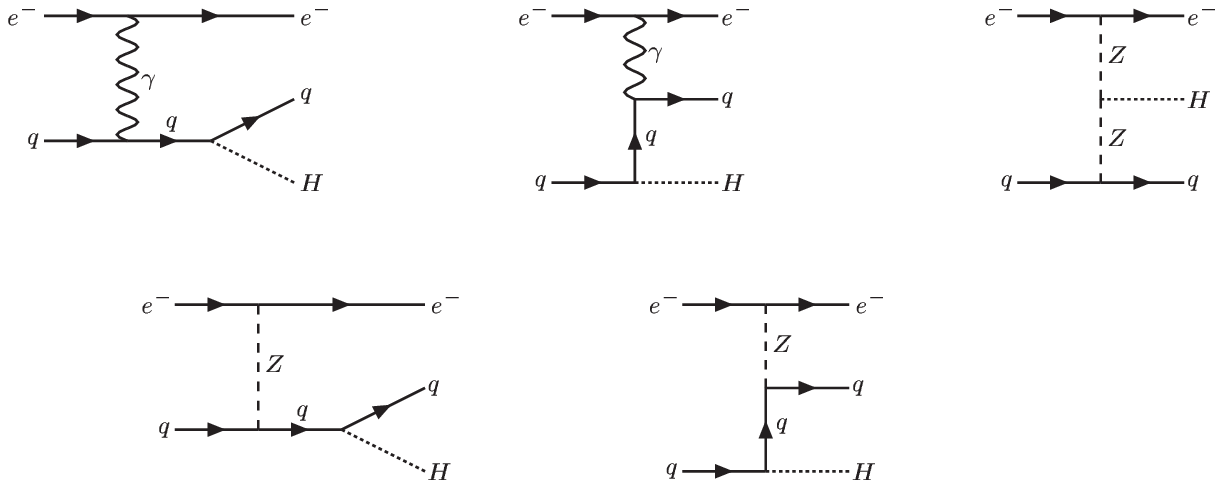}

\caption{Leading order diagrams for the subprocess $e^{-}q\rightarrow e^{-}Hq$
\label{fig.1}}
\end{figure}

\section{The Cross Section}

The calculation of the cross sections for signal and background is
performed using CalcHEP \cite{CalcHEP29} with parton distribution
function CTEQ6L \cite{PDF30}. For the background process $e^{-}p\to e^{-}HqX$,
we calculate the cross section values as $15.68$ fb and $39.75$
fb for the center of mass energies $\sqrt{s}=1.29$ TeV and $\sqrt{s}=1.98$
TeV, respectively. Here the anomalous couplings $\Delta a_{Z}$, $\beta_{Z}$
and $b_{Z}$ are taken to be zero for calculation of the cross sections
corresponding to the SM case. Although hadronic jets from $b$ quarks
would tag with high efficiency, lighter quarks tagging displays strong
rejections. We assume $b$ jet tagging efficiency of $\epsilon_{b}=60$\%
in the range $|\eta|<2.5$, apply the cuts on transverse momentum
of all final state particles of $p_{T}^{e,j}>30$ GeV and pseudorapidity
cut of $|\eta|<5$ on electron and jets. 

\begin{figure}
\includegraphics{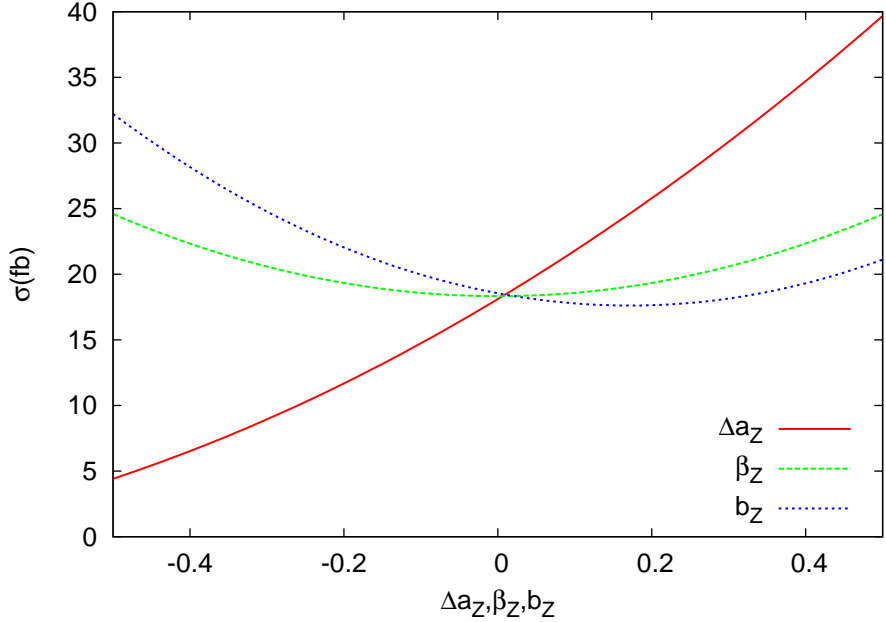}\caption{For the process $e^{-}p\rightarrow e^{-}HqX$, the dependence of the
cross section on anomalous couplings $\Delta a_{Z}$, $b_{Z}$ and
$\beta_{Z}$ for electron beam energy of 60 GeV \label{fig2}. }
\end{figure}

\begin{figure}
\includegraphics{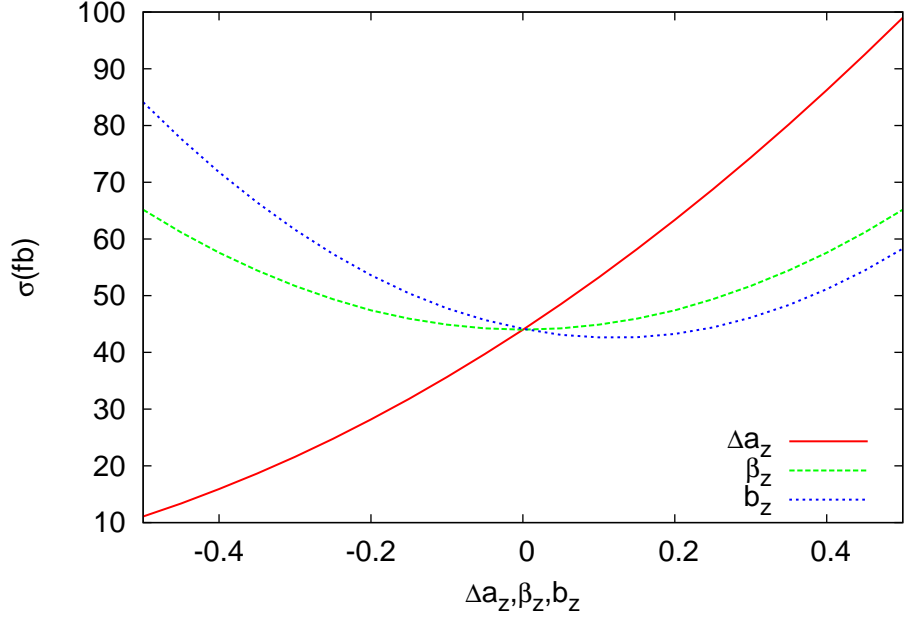}\caption{The same as \ref{fig2}, but for the electron beam energy of 140 GeV.\label{fig3} }
\end{figure}

Total cross section for $ep\to eHqX$ process as a function of anomalous
couplings $\Delta a_{Z},$$\,\beta_{Z}$ and $b_{Z}$ are given in
Fig. \ref{fig2} and Fig. \ref{fig3} for 60 and 140 GeV energies
of incoming electron, respectively. From these figures, we can see
a symmetric behavior around zero point for $\beta_{Z}$ and increasing
behaviour for $\Delta a_{Z}$, while the minimum of cross section
behaves shifted to about $0.2$ for $b_{Z}$.

\section{Constra\i{}nts on anomalous coupl\i{}ngs}

In this section, the sensitivity of the anomalous $HZZ$ couplings
is discussed. We estimate the sensitivity to these couplings at LHeC
for the integrated luminosities of $10$ and $100$ fb$^{-1}.$ We
use $\chi^{2}$ function to obtain sensitivity:

\begin{equation}
\chi^{2}=\left(\frac{\sigma_{SM}-\sigma_{A}}{\Delta\sigma_{SM}}\right)^{2}\label{eq:2}
\end{equation}
where $\sigma_{A}$ is the cross section including the anomalous couplings
$\Delta a_{Z}$, $b_{Z}$ and $\beta_{Z}.$ The error on the SM cross
section is defined as $\Delta\sigma_{SM}=\sigma_{SM}\delta_{stat.}$
with $\delta_{stat.}=1/\sqrt{N_{SM}}$; $N_{SM}$ is the number of
events within the SM calculated as $N_{SM}=\sigma_{SM}\times BR(H\rightarrow b\bar{b})\times(\epsilon_{b-tag})^{2}\times L_{int}$.
The SM cross section $(\sigma_{SM})$ corresponds to process $ep\rightarrow eHjX$
with vanishing anomalous couplings, where $j$ denotes light quarks. 

\begin{figure}
\includegraphics{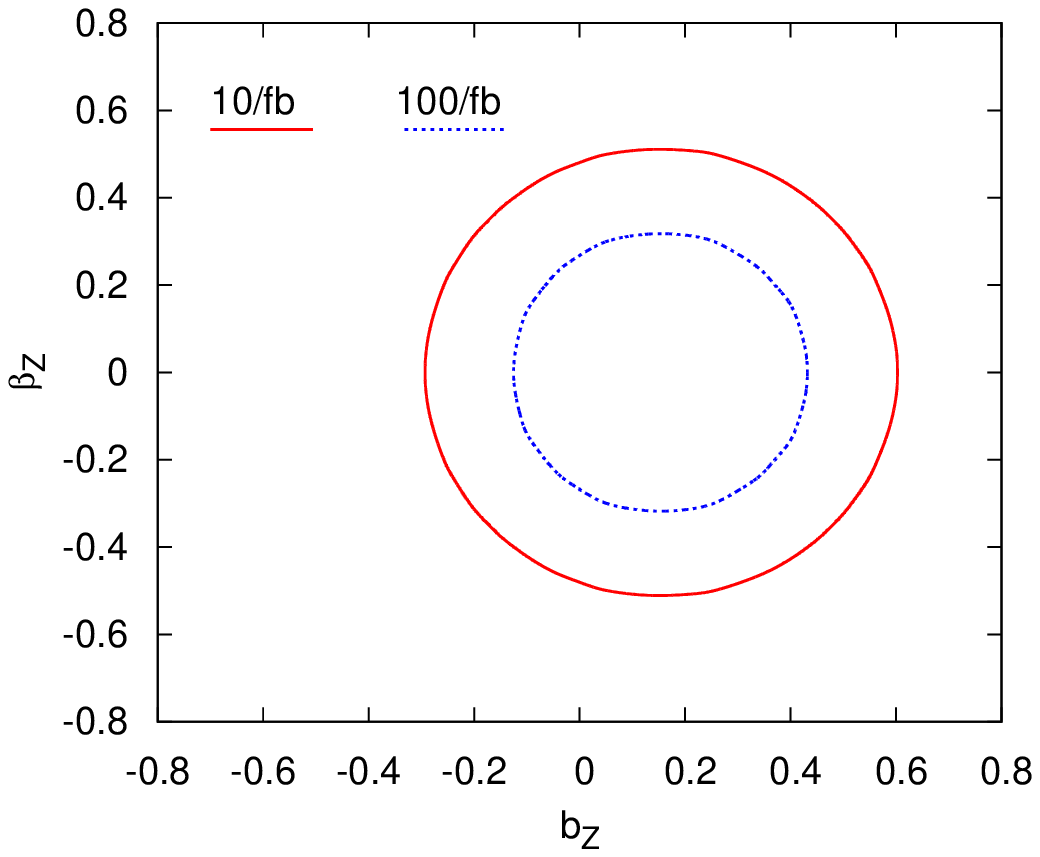}

\caption{Accessible region (with $95\%$ C.L.) of anomalous couplings at $E_{e}=60$
GeV. \label{fig4} }
\end{figure}

\begin{figure}
\includegraphics{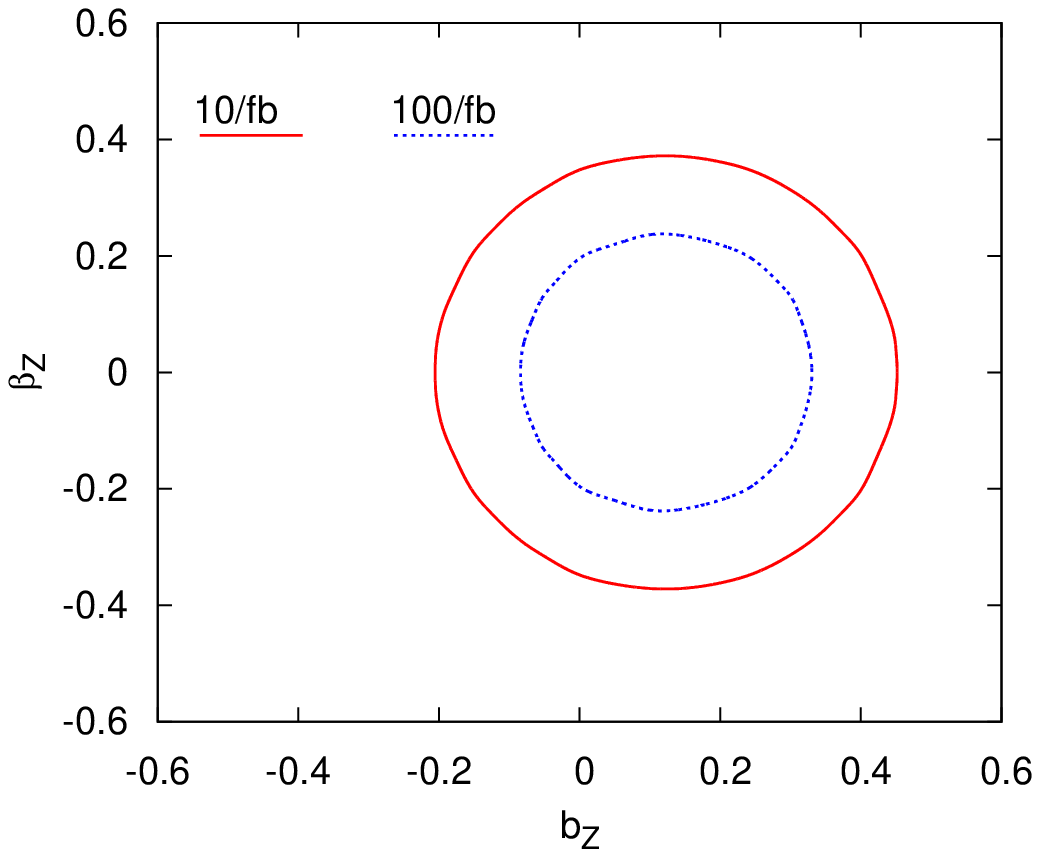}

\caption{Accessible region (with $95\%$ C.L.) of anomalous couplings at $E_{e}=140$
GeV. \label{fig5}}
\end{figure}

For an electron beam energy of $E_{e}=60$ GeV, the $95\%$ C.L. accessible
region in the plane $b_{Z}-\beta_{Z}$ is presented in Fig. \ref{fig4}
for the integrated luminosities 10 fb$^{-1}$ and 100 fb$^{-1}$.
As one can see, higher luminosity provides better limits on the anomalous
couplings. One can benefit from the higher center of mass energy corresponding
to $E_{e}=140$ GeV in order to obtain more sensitivity on the anomalous
couplings as shown in Fig. \ref{fig5}. The contours has an origin
at ($0.15,0.0$) for ($b_{Z},\beta_{Z}$), which is due to the cross
section behaviour as a function of the parameter $b_{Z}$ as shown
in Fig. \ref{fig2}. Assuming the SM value for $\triangle a_{Z}=0$
and $L_{int}=$10 fb$^{-1}$, the accesible limits for the anomalous
couplings are obtained as $-0.29<b_{Z}<0.61$ ($-0.51<\beta_{Z}<0.51$)
and $-0.21<b_{Z}<0.45$ ($-0.37<\beta_{Z}<0.37$) at $E_{e}=60$ GeV
and $E_{e}=140$ GeV, respectively. For an integrated luminosity of
$L_{int}=$100 fb$^{-1}$, we find the limits on the couplings $-0.12<b_{Z}<0.43$
($-0.32<\beta_{Z}<0.32$) and $-0.10<b_{Z}<0.33$ ($-0.24<\beta_{Z}<0.24$)
at $E_{e}=60$ GeV and $E_{e}=140$ GeV, respectively.

\section{Conclus\i{}ons}

The $e^{-}p\to e^{-}HqX$ process has smaller cross section when compared
to $e^{-}p\to\nu_{e}HqX$ process \cite{TauHan2010,Sudhansu2012,Senol2012},
but it has the advantage of electron identification according to the
missing neutrino in the latter case. We benefit from the kinematical
cuts explained in the previous section to reduce leading background
arising from the reactions including two $b$-jets in the final state.
We show that the LHeC at $\sqrt{s}=1.9$ TeV with luminosity of 100
$fb^{-1}$could allow the measurement of $HZZ$ anomalous couplings
$b_{Z}$ in the interval $(-0.10,\,0.33)$ and $\beta_{Z}$ in $(-0.24,\,0.24)$.
This result can be compared with the result from \cite{LHC} which
establishes a limit on anomalous coupling $\beta_{Z}=0.25$ at the
LHC with 14 TeV and $L=30$ $fb^{-1}$. Using the feasibility of measurement
of the bottom Yukawa coupling at the LHeC, it has the potential to
enhance the efficiency of the overall Higgs boson signal.

\end{document}